\documentclass[aps,prc,twocolumn,showpacs,superscriptaddress]{revtex4-1}
\usepackage{amsmath}
\usepackage{epsfig}
\usepackage{color}
\usepackage{endnotes}
\usepackage{natbib}
\usepackage[colorlinks,breaklinks]{hyperref}
\usepackage{nicefrac}
\usepackage{bm}
\usepackage{dcolumn}
\usepackage{graphics}
\usepackage{epstopdf}
\usepackage{amsmath}
\usepackage{amssymb}
\usepackage{color}
\usepackage{changes}
\newcommand{\bra}{\langle}
\newcommand{\ket}{\rangle}

\newcommand{\bs}[1]{\ensuremath{\boldsymbol{#1}}}

\newcommand{\be}{\begin{equation}}
\newcommand{\ee}{\end{equation}}
\newcommand{\bea}{\begin{align}}
\newcommand{\eea}{\end{align}}
\newcommand{\beqa}{\begin{eqnarray}}
\newcommand{\eeqa}{\end{eqnarray}}

\begin{document}

\title{Generalized nuclear contacts and the nucleon's momentum distributions}

\author{Ronen Weiss}
\affiliation{The Racah Institute of Physics, The Hebrew University, 
             Jerusalem, Israel}
\author{Betzalel Bazak}
\affiliation{The Racah Institute of Physics, The Hebrew University, 
             Jerusalem, Israel}
\author{Nir Barnea}
\email{nir@phys.huji.ac.il}
\affiliation{The Racah Institute of Physics, The Hebrew University, 
             Jerusalem, Israel}

\date{\today}

\begin{abstract} 
The general nuclear contact matrices are defined, taking into consideration
 all partial waves and finite-range interactions, extending Tan's work
for the zero range model. The properties 
of these matrices are discussed and the relations between the contacts and the
one-nucleon and two-nucleon momentum distributions are derived.
Using these relations, a new asymptotic connection between the one-nucleon
and two-nucleon momentum distributions, describing the two-body short-range
correlations in nuclei, is obtained.
Using available numerical data, we extract few connections between
the different contacts and verify their relations to the momentum distributions.
The numerical data also allows us to identify the main nucleon momentum range
affected by two-body short-range correlations. Utilizing these relations
and the numerical data, we also verify a previous independent prediction
connecting between the Levinger constant and the contacts.
This work provides an important indication for the relevance of the contact formalism to 
nuclear systems, and should open the path for revealing more useful relations
between the contacts and interesting quantities of nuclei and nuclear matter.
\end{abstract}


\maketitle
\section{Introduction}
Recently, a new variable, called \emph{the contact}, was defined by Tan 
\cite{Tan08,Bra12} for a two component Fermi gas interacting via 
short-range forces.
The contact measures the probability to find
two unlike fermions close to each other. 
In a series of theorems, called Tan's relations, 
many other properties of the system, such as its energy, pressure, and momentum 
distribution, were connected to the contact. 
Tan assumes that the range of the interaction is much smaller than
the scattering length and the averaged distance between the fermions.
Following these theoretical predictions, several experiments 
were conducted, verifying Tan's relations in ultracold atomic systems
consisting of $^{40}$K \cite{SteGaeDra10,SagDraPau12}
and $^6$Li \cite{ParStrKam05,WerTarCas09,KuhHuLiu10} atoms.

Nuclear systems differ from these ultracold atomic systems
in many aspects. First, the nucleons are not two-component
fermions. Second, while in the atomic systems the strength 
of the interaction between the atoms and the density can
be changed easily, such that Tan's assumptions are satisfied,
in nuclear physics it cannot be done. In nuclear systems,
the $s$-wave spin-singlet and spin-triplet scattering lengths are about
$-20$ fm and $5.38$ fm, respectively, and the average distance between
two adjacent nucleons is about $2.4$ fm. The interaction
range of the long range part of the nuclear potential, which is
governed by the pion exchange Yukawa force, is about
$\mu^{-1}=\hbar/m_\pi c \approx 1.4$ fm. Thus, in nuclear physics 
the interaction range is only slightly smaller than the average distance 
between two particles and the scattering length.
Consequently, some changes are to be done in order to generalize Tan's 
relations to nuclear systems.

Considering a two-component Fermi gas that obeys Tan's assumptions,
the high momentum tail of the momentum distribution is connected to the contact
through the relation, $n_{\sigma}(k)\rightarrow C/k^4$ as $k\rightarrow\infty$, 
where $n_\sigma (k)$ is the momentum distribution of fermions with spin
$\sigma$, and $C$ is the contact. 
In nuclear physics, the high-momentum part of the nucleon's
momentum distribution is one of the main tools for studying short range correlations (SRCs)
between nucleons. The main focus in current studies
of two-body SRCs (see e.g. 
\cite{WirSchPie14,Fmoin12,HenSci14,AlvCioMor08,ArrHigRos12})
 is around the momentum range
$1.5\: \mathrm{fm^{-1}} < k < 3 \: \mathrm{fm^{-1}}$.
In few of these studies it is claimed that
higher momentum is affected also by
3-body correlations \cite{Egiyan06}. In this momentum range,
a dominance of neutron-proton ($np$) correlated pairs was observed
in electron scattering experiments \cite{HenSci14}. 
This $np$ dominance is usually explained by
the contribution of the tensor force, which affects
only spin-triplet $np$ pairs.
Another 
observation is that the correlated pairs usually have
high relative momentum and low center of mass momentum, i.e. they
move approximatelly back-to-back.
Generalizing Tan's relation between the high momentum tail
and the contact to nuclear systems, should help in understanding
more properties of SRCs in nuclei.

In a previous paper \cite{WeiBazBar14}, we have suggested that it might be 
fruitful to use the contact formalism in nuclear systems. There we have defined 
the neutron-proton $s$-wave nuclear contacts and evaluated their average value 
by relating them to the Levinger constant of the photoabsorption process. 
In this work we 
generalize the definition of the nuclear contacts from $s$-wave
to all partial waves. We also consider finite-range interactions instead
of zero-range. The result is the matrices of nuclear contacts. 
We discuss the properties of these matrices, and 
use our generalized contact formalism to relate
the nuclear contacts to the one-nucleon and two-nucleon momentum distributions
in nuclei. Doing so, we find an asymptotic relation between these two 
distributions which is relevant to the study of SRCs in nuclei. 
This relation is verified by available numerical data. 
Further analysis of the numerical data
and its implications to the contact formalism are also presented.

In this paper we focus on the two-body contacts
and on two-body correlations, postponing the discussion on
three-body effects to future publications.

\section{The matrices of nuclear contacts}

Consider a two-component Fermi gas that obeys Tan's assumptions.
In such a gas, when a spin-up particle $i$ gets close to a spin-down
particle $j$, the many-body wave function can be factorized into a 
product of an asymptotic pair wave function $\varphi(\bs r_{ij})$,
$\bs r_{ij}=\bs{r}_i-\bs{r}_j$,
and a function $A$, also called the regular part of $\Psi$,
describing the residual $A-2$ particle system
and the pair's center of mass $\bs{R}_{ij}=(\bs{r}_i+\bs{r}_j)/2$
motion \cite{Tan08,WerCas12},
\be \label{wf}
  \Psi \xrightarrow[r_{ij}\rightarrow 0]{}\varphi(\bs{r}_{ij})
           A(\bs{R}_{ij},\{\bs r_k\}_{k\neq i,j})\;.
\ee
Due to the suppression of higher partial waves in these systems,
the asymptotic pair wave function will be predominantly an $s$-wave.
In particular, in the zero-range model \cite{zerorange} the
pair wave function is given by
$\varphi=\left(1/r_{ij}-1/a\right)$, where $a$ is
the scattering length.

The contact $C$ is then defined by \cite{Tan08,WerCas12}
\be\label{contact_generic}
   C=16\pi^2 N_{\uparrow\downarrow} \bra A| A\ket,
\ee
where 
\begin{align}
   \bra A|  A\ket & = 
    \int  \prod_{k\neq i,j} d\bs{r}_{k} \,d\bs{R}_{ij} \,
    \\ \nonumber & \times
        A^{\dagger}\left(\bs{R}_{ij},\{\bs{r}_{k}\}_{k\neq i,j}\right)
        \cdot
        A\left(\bs{R}_{ij},\{\bs{r}_{k}\}_{k\neq i,j}\right)\;
\end{align}
and $N_{\uparrow\downarrow}$ is the number of possible spin up - spin down pairs.

In nuclear physics, we have four-component fermions, 
which are the protons and neutrons with their spin being either up or down.
Moreover, the assumption of a
zero-range $s$-wave interaction is not accurate for nuclei.
As a result, few changes must be made in order to generalize the contact
formalism to study nuclear systems.

A nucleus can be described by a wave function $\Psi$ with 
total angular momentum $J$ and projection $M$. We will assume that when
particle $i$ gets close to particle $j$, the wave function is
still factorized but the pair wave function depends on the total
spin of the pair $s_2$, and its angular momentum quantum number $\ell_2$ 
(with
respect to the relative coordinate $\bs{r}_{ij}$) which
are coupled to create the total pair angular momentum $j_2$
and projection $m_2$.
The asymptotic form of the wave function is then given by
\be\label{full_asymp}
\Psi\xrightarrow[r_{ij}\rightarrow 0]{}\sum_\alpha\varphi_{ij}^\alpha\left(\bs{r}_{ij})A_{ij}^\alpha(\bs{R}_{ij},\{\bs{r}_k\}_{k\not=i,j}\right).
\ee
Here the index $ij$ corresponds to one of 
the three particle pairs: proton-proton ($pp$),
neutron-neutron ($nn$) or neutron-proton ($np$). We note that
due to symmetry the asymptotic functions are invariant under same particle
permutations.
The sum over $\alpha$ denotes a sum over
the four quantum numbers $\left(s_2,\ell_2,j_2,m_2\right)$.
\be
A_{ij}^\alpha=\sum_{J_{A-2},M_{A-2}}\bra j_2m_2J_{A-2}M_{A-2}|JM\ket A_{ij}^{\{s_2,\ell_2,j_2\}J_{A-2},M_{A-2}}\;.
\ee
Here, $J_{A-2}$ and $M_{A-2}$ are the angular momentum
quantum numbers with respect to
 $\bs{J}_{A-2}+\bs{L}_{2,CM}$, where $\bs{J}_{A-2}$ is the total angular
 momentum of the residual $(A-2)$ particles 
and $\bs{L}_{2,CM}$ is the spatial angular momentum with respect to $\bs{R}_{ij}$.
$A_{ij}^{\{s_2,\ell_2,j_2\}J_{A-2},M_{A-2}}$ is a set of functions
with angular momentum quantum numbers $J_{A-2}$ and $M_{A-2}$,
which depends also on the numbers $s_2,\ell_2,j_2$.
\be
\varphi_{ij}^{\alpha}\equiv\varphi_{ij}^{(\ell_2s_2)j_2m_2}=
 [\varphi_{ij}^{\{s_2,j_2\}\ell_2}\otimes\chi_{s_2}]^{j_2m_2}
\;,
\ee 
where $\chi_{s_2\mu_s}$ is the two-body spin function, and 
$\varphi_{ij}^{\{s_2,j_2\}\ell_2\mu_\ell}(\bs{r}_{ij})
    ={\phi}_{ij}^{\{\ell_2,s_2,j_2\}} (r_{ij}) Y_{\ell_2 \mu_\ell}(\hat{r}_{ij})$.
For clarity, when angular momentum indices are written without any
brackets they denote the relevant angular momentum quantum numbers of the 
function.
When the indices are in curly brackets, it means that the function depends 
on this numbers but they do not denote the angular momentum of the function.
When two indices are inside round brackets, it means that the angular momentum
of the function is created by a coupling of these two indices.

The only assumption we make regarding the set of functions 
$\{\varphi_{ij}^\alpha\}$
is that they do not depend on the specific nuclei or its total angular momentum
$J$ and $M$. 
This is a reasonable assumption, because when two particles are very close they
interact with 
each other regardless to the background of the $A-2$ particle system. Doing so, 
we no longer use the $s$-wave or the zero-range assumptions.

Since the $A_{ij}^\alpha$ functions are not generally orthogonal for different $\alpha$, we are led 
to define matrices of nuclear contacts in the following way:
\be\label{JM_contacts_def}
C_{ij}^{\alpha \beta}(JM)=16{\pi}^2N_{ij}\bra A_{ij}^\alpha | A_{ij}^\beta \ket.
\ee
As before, $ij$ stands for one of the pairs: $pp$, $nn$ or $np$, 
$N_{ij}$ is the number of $ij$ pairs, and $\alpha$ and $\beta$ are the matrix
 indices.
We also denote
$\alpha=(s_\alpha,\ell_\alpha,j_\alpha,m_\alpha)$
and $\beta=(s_\beta,\ell_\beta,j_\beta,m_\beta)$. One can see that if
$m_\alpha\neq m_\beta$, then $C_{ij}^{\alpha \beta}(JM)=0$, but it is not generally true
for $j_2$, $s_2$ or $\ell_2$. For spherical nuclei $(J=0)$ we do get 
$C_{ij}^{\alpha \beta}(JM)=0$ if $j_\alpha \neq j_\beta$. For $pp$ and $nn$ pairs, 
Pauli's exclusion principle tells us that unless $s_\alpha+\ell_\alpha$ is even, 
we have $A_{pp}^\alpha=A_{nn}^\alpha=0$, so $C_{pp}^{\alpha\beta}=C_{nn}^{\alpha\beta}=0$ 
if $s_\alpha+\ell_\alpha$ or $s_\beta+\ell_\beta$ are odd.
Moreover, if $\Psi$ is the ground state of the nucleus, or any eigenstate of the nuclear Hamiltonian,
then $\Psi$ has a defined parity. $\varphi_{ij}^\alpha$ has a parity of $(-1)^{\ell_\alpha}$, so 
it dictates the parity of $A_{ij}^\alpha$. Thus, $C_{ij}^{\alpha\beta}(JM)=0$ for $\alpha$
and $\beta$ such that $\ell_\alpha$ and $\ell_\beta$ have different parities.

Since the projection $M$ is usually unknown in experiments, it is useful 
to define the averaged nuclear contacts:
\be\label{ave_contacts_def}
C_{ij}^{\alpha \beta}=\frac{1}{2J+1}\sum_M C_{ij}^{\alpha\beta}(JM).
\ee
According to this definition, we have three matrices of averaged contacts,
one for each kind of nucleon-nucleon pair.
We note that the averaged contacts still depend on
$J$, but we will not write it explictly.
Using Clebsch Gordan identities
one can prove that if $m_\alpha\neq m_\beta$
or $j_\alpha \neq j_\beta$, then $C_{ij}^{\alpha \beta}=0$, and 
also that the averaged contacts are independent of $m_\alpha$ and $m_\beta$.
The averaged contacts inherit the properties of the non-averaged contacts 
$C_{ij}^{\alpha\beta}(JM)$
regarding parity and Pauli's principle.

Concluding, for a given $\alpha$, the relevant $\beta$'s such that 
$C_{ij}^{\alpha\beta}$
can be different from zero must obey $j_\beta=j_\alpha$ and $m_\beta=m_\alpha$.
Since, $s_2=0,1$ there are four $(s_2,\ell_2)$ pairs that can create a given $j_\alpha\neq 0$:
$(0,j_\alpha)$, $(1,j_\alpha)$, $(1,j_\alpha-1)$ and $(1,j_\alpha+1)$. The first two
options have the same parity of $\ell_2$ and the last two have the opposite 
parity. For $j_\alpha=0$ we have only two possible $(s_2,\ell_2)$ pairs:
$(0,0)$ and $(1,1)$, which have different parity of $\ell_2$.
Thus, in general the matrices $C_{ij}^{\alpha\beta}$ are built from $2\times2$ blocks,
except for the two $1\times 1$ blocks associated with the $j_2=0$ case.
Each block has a well defined $j_2,m_2$ values. For any $j_2\neq 0$
there are two blocks, one with $(s_2,\ell_2)=(0,j_2),(1,j_2)$
and the other with $(s_2,\ell_2)=(1,j_2-1),(1,j_2+1)$.
For $pp$ and $nn$ pairs, Pauli's principle dictates
that any matrix element with an odd
$s_2+\ell_2$ is zero, so some of the $2 \times 2$ blocks
are reduced into two $1 \times 1$ blocks.

In a previous paper \cite{WeiBazBar14}
we have defined the $s$-wave nuclear contacts, $C_{ij}^{s_2}(JM)$, for $s_2=0,1$.
The definition there was slightly different from the current one, as the two-body
spin functions were included into the regular $(A-2)$ particle function
$A^{\alpha}_{ij}$. In our current definition,
we have four diagonal $s$-wave contacts $C_{ij}^{\alpha_{00}\alpha_{00}}$ 
and
$C_{ij}^{\alpha_{1\mu}\alpha_{1\mu}}$, where
$\alpha_{00}=(s_2=0,\ell_2=0,j_2=0,m_2=0)$, 
$\alpha_{1\mu}=(s_2=1,\ell_2=0,j_2=1,m_2=\mu)$, and $\mu=-1,0,1$.
The relations between the two definitions are
\be 
C_{ij}^{s_2=0}(JM)=C_{ij}^{\alpha_{00}\alpha_{00}}(JM)
\ee
\be
C_{ij}^{s_2=1}(JM)=\sum_{\mu=-1}^{1}C_{ij}^{\alpha_{1\mu}\alpha_{1\mu}}(JM)
\ee
Averaging over $M$ and using the fact 
that the averaged contacts are independent of $m_2$
we get
\be \label{rel_singlet}
C_{ij}^{s_2=0}=C_{ij}^{\alpha_{00}\alpha_{00}}
\ee
\be \label{rel_triplet}
C_{ij}^{s_2=1}=\sum_{\mu=-1}^{1}C_{ij}^{\alpha_{1\mu}\alpha_{1\mu}}=3C_{ij}^{\alpha_{1\mu}\alpha_{1\mu}}
\ee
We also note that the previously defined
$s$-wave contacts, $C_{ij}^{s_2}(JM)$, are actually independent of $M$.
Thus, also $C_{ij}^{\alpha_{00}\alpha_{00}}(JM)$
and $\sum_{\mu=-1}^{1}C_{ij}^{\alpha_{1\mu}\alpha_{1\mu}}(JM)$
are independent of $M$.

\section{Momentum distributions}
\subsection{The two-nucleon momentum distribution}

In the following we will utilize the above generalized contact formalism 
to find a relation between the two-nucleon momentum distribution 
and the nuclear contacts.

Let's denote by $f_{ij}^{JM}(\bs{k}+\bs{K}/2, -\bs{k}+\bs{K}/2)$ the
density probability to find a pair of nucleons, $ij\in\{pp,nn,pn\}$, with 
any particle of type $i$ 
with momentum $\bs{k}+\bs{K}/2$ and any particle of type $j$ 
with momentum $-\bs{k}+\bs{K}/2$. $J$ and $M$
are the angular momentum quantum numbers of the nuclear wave function
 $\Psi$.
Working in the momentum space
\be
\tilde{\Psi}(\bs{k}_1,...,\bs{k}_A)=\int  \prod_{n=1}^A d^3r_n \Psi e^{\sum_{n}i\bs{k}_n\cdot\bs{r}_n},
\ee
and we can write
\begin{align}\label{fij}
& f_{ij}^{JM}(\bs{k}+\bs{K}/2, -\bs{k}+\bs{K}/2)=
      N_{ij} \int  \prod_{m\neq i,j} \frac{d^3k_m}{(2\pi)^3} \nonumber \\
 & \times    
\left| \tilde{\Psi}(\bs{k}_1,...,\bs{k}_i=\bs{k}+\bs{K}/2,...
                   ,\bs{k}_j=-\bs{k}+\bs{K}/2,...,\bs{k}_A)\right| ^2
\end{align}
where A is the number of nucleons, $N_{ij}$ is the number of $ij$ pairs, and we notice that $f_{ij}^{JM}$
is normalized in such away that 
$\int f_{ij}^{JM} \frac{d^3k}{(2\pi)^3} \frac{d^3K}{(2\pi)^3}=N_{ij}$.

In the limit
$k\rightarrow \infty$ the main contribution to $f_{ij}^{JM}$
comes from the asymptotic $\bs{r}_{ij}\rightarrow 0$ part of the wave function,
given in Eq. (\ref{full_asymp}). All other terms will cancel each other
due to the 
fast oscillating $\exp(i\bs{k}\cdot \bs{r}_{ij})$ factor.
Substituting $\tilde{\Psi}$ into Eq. (\ref{fij}), and using Eq. (\ref{full_asymp})
we get
\begin{align}\label{fij_asymp}
& f_{ij}^{JM}(\bs{k}+\bs{K}/2, -\bs{k}+\bs{K}/2)=
    N_{ij} \int  \prod_{m\neq i,j} \frac{d^3k_m}{(2\pi)^3} \nonumber \\
 & \times    
| \int  \prod_{n\neq i,j}^A d^3r_nd^3r_{ij}d^3R_{ij} \sum_\alpha \varphi_{ij}^\alpha(\bs{r}_{ij})A_{ij}^\alpha 
 \nonumber \\ 
& \times 
\exp({i\bs{k} \cdot \bs{r}_{ij}}+{i\bs{K}\cdot \bs{R}_{ij}}+{\sum_{n\neq i,j}i\bs{k}_n\cdot\bs{r}_n})
| ^2.
\end{align}
We will define now
\be
F_{ij}^{JM}(\bs{k})=\int \frac{d^3K}{(2\pi)^3}  f_{ij}^{JM}(\bs{k}+\bs{K}/2, -\bs{k}+\bs{K}/2).
\ee
$F_{ij}^{JM}$ is the density probability to find an $ij$ pair with relative momentum $\bs{k}$, and
it obeys the normalization condition $\int F_{ij}^{JM}(\bs{k})\frac{d^3k}{(2\pi)^3}=N_{ij}$. 
We can now substitute the asymptotic form of $f_{ij}^{JM}$, Eq. (\ref{fij_asymp}), into the definition
of $F_{ij}^{JM}$. In the resulting expression we can separate the integration over
 $\bs{r}_{ij}$ from the rest of the coordinates.  
Using the notation 
\be
\tilde{\varphi}_{ij}^\alpha(\bs{k})=\int d^3r \varphi_{ij}^\alpha(\bs{r})\exp(i\bs{k}\cdot\bs{r})
\ee
and 
\be
\tilde{A}_{ij}^\alpha=\int \prod_{n \neq i,j}d^3r_nd^3R_{ij}A_{ij}^\alpha 
        \exp(i\bs{K}\cdot \bs{R}_{ij}+\sum_{n \neq i,j} i\bs{k}_n\cdot \bs{r}_n)\;,
\ee
we get
\begin{align}
&F_{ij}^{JM}(\bs{k})= \nonumber\\
& 
N_{ij}\sum_{\alpha,\beta}\tilde{\varphi}_{ij}^{\alpha\dagger} (\bs{k}) \tilde{\varphi}_{ij}^\beta (\bs{k})
      \int \prod_{m \neq i,j} \frac{d^3k_m}{(2\pi)^3}\frac{d^3K}{(2\pi)^3}
      \tilde{A}_{ij}^{\alpha\dagger}\tilde{A}_{ij}^\beta.
\end{align}
Noting the equality
\be
 \int \prod_{m \neq i,j} \frac{d^3k_m}{(2\pi)^3}\frac{d^3K}{(2\pi)^3}
    \tilde{A}_{ij}^{\alpha\dagger}\tilde{A}_{ij}^\beta
    =\int \prod_{n \neq i,j}d^3r_n d^3R_{ij}A_{ij}^{\alpha\dagger}A_{ij}^\beta \;,
\ee
we obtain the following asymptotic $k\rightarrow \infty$ expression for the two nucleon
momentum distribution,
\be
F_{ij}^{JM}(\bs{k})=\sum_{\alpha,\beta}
   \tilde{\varphi}_{ij}^{\alpha\dagger}(\bs{k}) \tilde{\varphi}_{ij}^\beta (\bs{k})
   \frac{C_{ij}^{\alpha\beta}(JM)}{16\pi^2}.
\ee
Here we have used the definition of the 
contacts from Eq. (\ref{JM_contacts_def}). Averaging over $M$, we get
the asymptotic relation 
\be \label{2nuc}
F_{ij}(\bs{k})=\sum_{\alpha,\beta}
      \tilde{\varphi}_{ij}^{\alpha\dagger}(\bs{k})\tilde{\varphi}_{ij}^\beta(\bs{k})
      \frac{C_{ij}^{\alpha\beta}}{16\pi^2},
\ee
where $F_{ij}=(2J+1)^{-1}\sum_M F_{ij}^{JM}$, and 
$C_{ij}^{\alpha\beta}$ are the averaged contacts defined in Eq. (\ref{ave_contacts_def}).
Like $C_{ij}^{\alpha\beta}$, also $F_{ij}$ depends implicitly on $J$.

 \subsection{The one-nucleon momentum distribution}

We would like now to connect the nuclear contacts also to the 
one-nucleon momentum distributions. The following derivation is
based on Tan's derivation for the two-body case in atomic systems \cite{Tan08}.
We will first focus on the proton's momentum distribution $n_p^{JM}(\bs{k})$.
Normalized to the number of protons in the system $Z$,
 $\int\frac{d^3k}{(2\pi)^3}n_p^{JM}(\bs{k})=Z$,
$n_p^{JM}$ is given by
\begin{align}\label{n_p_jm}
&n_p^{JM}(\bs{k})=Z \int \prod_{l \neq p} \frac{d^3k_l}{(2\pi)^3} 
\left| \tilde{\Psi}(\bs{k}_{1},...,\bs{k}_{p}=\bs{k},...,\bs{k}_{A})\right|^2,
\end{align}
where $p$ is any proton.

In the $k\longrightarrow \infty$ limit the main contribution to $n_p^{JM}$
emerges from the asymptotic parts of the wave function, i.e.
from $r_{p s}=|\bs{r}_{p}-\bs{r}_s|\rightarrow 0$, for any particle $s \neq p$, being proton or neutron.
In this limit
\begin{align} 
\tilde{\Psi}(\bs{k}_{1},...,&\bs{k}_{p}=\bs{k},...,\bs{k}_{A})= 
 \sum_{s \neq p} \sum_\alpha\tilde{\varphi}_{p s}^\alpha\left((\bs{k}-\bs{k_s})/2\right) 
\nonumber \\  & \times   
      \tilde{A}_{p s}^\alpha\left(\bs{K}_{p s}=\bs{k}+\bs{k}_s,\{\bs{k}_j\}_{j \neq p,s}\right),
\end{align}
where $\bs{K}_{p s}$ is the center of mass momentum of the $p s$ pair.
Substituting into $n_p^{JM}(\bs{k})$, we see that 
since $A_{p s}^\alpha$ is regular, $\tilde{A}_{p s}^\alpha$
will be significant only if $\left| \bs{k}+\bs{k}_s \right| \ll k$. 
It means that $\bs{k}_s\approx -\bs{k}$ so $\bs{k}-\bs{k}_s\approx 2\bs{k}$.
Substituting $\tilde\Psi^\dagger\tilde\Psi$ into Eq. (\ref{n_p_jm}), we get summations over 
$s,s' \neq p$.
The contribution of the $s, s'$ element, for $s \neq s'$,
will be significant only for $\bs{k}_s\approx \bs{k}_{s'} \approx -\bs{k}$.
In this case $k,k_s,k_{s'}\rightarrow \infty$ together, which
is clearly a three body effect and we expect it to be less important \cite{BraKanPla11}.
Therefore we are left only with the diagonal elements and get
\begin{align}
  n_p^{JM}(\bs{k})&= Z \sum_{s \neq p} \sum_{\alpha,\beta} 
     \int \prod_{l \neq p, s} \frac{d^3k_l}{(2\pi)^3}\frac{d^3K_{p s}}{(2\pi)^3}
    \tilde{\varphi}_{p s}^{\alpha\dagger}(\bs{k}) \tilde{\varphi}_{p s}^\beta(\bs{k}) \nonumber  \\
& \times
   \tilde{A}_{p s}^{\alpha\dagger}(\bs{K}_{p s},\{\bs{k}_j\}_{j \neq p,s}) 
   \tilde{A}_{p s}^\beta(\bs{K}_{p s},\{\bs{k}_j\}_{j \neq p,s}).
\end{align}
We will now divide the sum 
$\sum_{s \neq p}$ into a sum over protons and a sum
over neutrons $\sum_{p' \neq p}+\sum_n$. Since the asymptotic
functions $A_{pp'}^\alpha$ and $\varphi_{pp'}^\alpha$ are the same for all $pp'$ 
pairs we can take them out of the sum.
The same holds for the $np$ pairs. As a result we get
\begin{align}
    n_p^{JM}(\bs{k})&=\sum_{\alpha,\beta} \tilde{\varphi}_{pp}^{\alpha\dagger}(\bs{k}) 
                                      \tilde{\varphi}_{pp}^\beta(\bs{k})
         Z(Z-1) \langle A_{pp}^\alpha | A_{pp}^\beta \rangle \nonumber \\
& \times
\sum_{\alpha,\beta} \tilde{\varphi}_{pn}^{\alpha\dagger}(\bs{k}) \tilde{\varphi}_{pn}^\beta(\bs{k})
         NZ \langle A_{pn}^\alpha | A_{pn}^\beta \rangle.
\end{align}
Here $N$ is the number of neutrons in the system.
Using the definition of the contacts, Eq. (\ref{JM_contacts_def}),
we see that for $k\rightarrow \infty$
\begin{align}
  n_p^{JM}(\bs{k})&=\sum_{\alpha,\beta} 
  \tilde{\varphi}_{pp}^{\alpha\dagger}(\bs{k}) \tilde{\varphi}_{pp}^\beta(\bs{k}) 
  \frac{2C_{pp}^{\alpha\beta}(JM)}{16\pi^2}\nonumber \\
&
    +\sum_{\alpha,\beta} 
  \tilde{\varphi}_{pn}^{\alpha\dagger}(\bs{k}) \tilde{\varphi}_{pn}^\beta(\bs{k})
  \frac{C_{pn}^{\alpha\beta}(JM)}{16\pi^2}.
\end{align}
Averaging over $M$ we further obtain the relation
between the averaged contacts and the averaged protons' momentum 
distribution $n_p(\bs{k})=(2J+1)^{-1}\sum_M n_p^{JM}(\bs{k})$ for $k\rightarrow\infty$:
\begin{align} \label{1p}
n_p(\bs{k})&=\sum_{\alpha,\beta} 
    \tilde{\varphi}_{pp}^{\alpha\dagger}(\bs{k}) \tilde{\varphi}_{pp}^\beta(\bs{k}) 
    \frac{2C_{pp}^{\alpha\beta}}{16\pi^2}
  \nonumber \\
& 
   +\sum_{\alpha,\beta} 
   \tilde{\varphi}_{pn}^{\alpha\dagger}(\bs{k}) \tilde{\varphi}_{pn}^\beta(\bs{k})
   \frac{C_{pn}^{\alpha\beta}}{16\pi^2}.
\end{align}
We note that $n_p(\bs{k})$ still depends on $J$.
Similarly, for the neutrons:
\begin{align} \label{1n}
n_n(\bs{k})&=\sum_{\alpha,\beta} \tilde{\varphi}_{nn}^{\alpha\dagger}(\bs{k}) \tilde{\varphi}_{nn}^\beta(\bs{k}) \frac{2C_{nn}^{\alpha\beta}}{16\pi^2} \nonumber \\
& 
+\sum_{\alpha,\beta} \tilde{\varphi}_{pn}^{\alpha\dagger}(\bs{k}) \tilde{\varphi}_{pn}^\beta(\bs{k})\frac{C_{pn}^{\alpha\beta}}{16\pi^2}.
\end{align}
Comparing Eqs. (\ref{1p}) and (\ref{1n}) to Eq. (\ref{2nuc}),
we can see that for $k\longrightarrow \infty$ there is a simple
relation between the one-nucleon and 
the two-nucleon momentum distributions:
\be \label{1pto2}
n_p(\bs{k})=2F_{pp}(\bs{k})+F_{pn}(\bs{k})
\ee
\be \label{1nto2}
n_n(\bs{k})=2F_{nn}(\bs{k})+F_{pn}(\bs{k}).
\ee
These connections seem intuitive if we assume 
that a nucleon will have high momentum $\bs{k}$ only if 
there is another nucleon close to it with opposite momentum
$-\bs{k}$. In this case, if we find a proton with high momentum 
$\bs{k}$ we know that we will find close to it a neutron or 
a proton, that is a correlated $pp$ or $np$ 
pair with relative momentum $\bs{k}$. Notice that the
factor of $2$ before $F_{pp}$ and $F_{nn}$ in Eqs.
(\ref{1pto2}), (\ref{1nto2}), can be also explained
in this picture by the fact that
for example a $pp$ pair with momenta $(-\bs{k},\bs{k})$ has 
a relative momentum $-\bs{k}$ even though there is a proton 
with momentum $\bs{k}$ in this pair. It means that such a pair
will be counted for $n_p(\bs{k})$ but not for $F_{pp}(\bs{k})$ and the 
factor of 2 takes it into consideration. 

These relations emphasize
the importance of the two-body correlations to the high momentum
one-nucleon distribution. As mentioned before,
the picture of short-range correlated pairs 
of nucleons with back-to-back momentum is one of the main features
of SRCs in nuclei, and the above relations between
the one-nucleon and two-nucleon momentum distributions give a 
theoretical support to this picture.

We also note here that similar derivations can be done
easily for atomic systems consisting of two-component fermions,
denoted by $\uparrow$ and $\downarrow$.
The one-body high momentum distribution is already known
and given by $n_{\uparrow}(\bs{k})=n_{\downarrow}(\bs{k})=C/k^4$.
Adjusting the above derivation for the two-nucleon momentum
distribution to atomic systems will produce an identical relation between
the two-body momentum distribution,
$F_{\uparrow\downarrow}(\bs{k})$, describing the 
probability to find an $\uparrow\downarrow$
pair with high relative momentum, and the atomic contact. Explicitly,
$F_{\uparrow\downarrow}(\bs{k})=C/k^4$. As a result we find that
$n_\uparrow(\bs{k})=n_\downarrow(\bs{k})=F_{\uparrow\downarrow}(\bs{k})$
for high momentum $\bs{k}$.
This relation tells us that also in the ultracold atomic systems
the correlated $\uparrow\downarrow$ pairs have
back-to-back momentum, like in nuclear systems.

\section{Analysis of numerical data}
\subsection{Momentum distributions}
In order to check the validity of our results in actual nuclear systems, 
we turn now to compare our theoretical predictions to  
available numerical data.
To this end, we will use numerical data of one-nucleon and
two-nucleon momentum distributions calculated by Wiringa et al. \cite{WirSchPie14},
using the Variational Monte Carlo method (VMC), for nuclei with $A\le 12$.
In this VMC results, the calculation of both one-nucleon and two-nucleon momentum
distributions were done for nuclei in their ground state.
Consequently, the following analysis is limited to the nuclear ground 
state.

\begin{figure}
\includegraphics[width=8.6 cm]{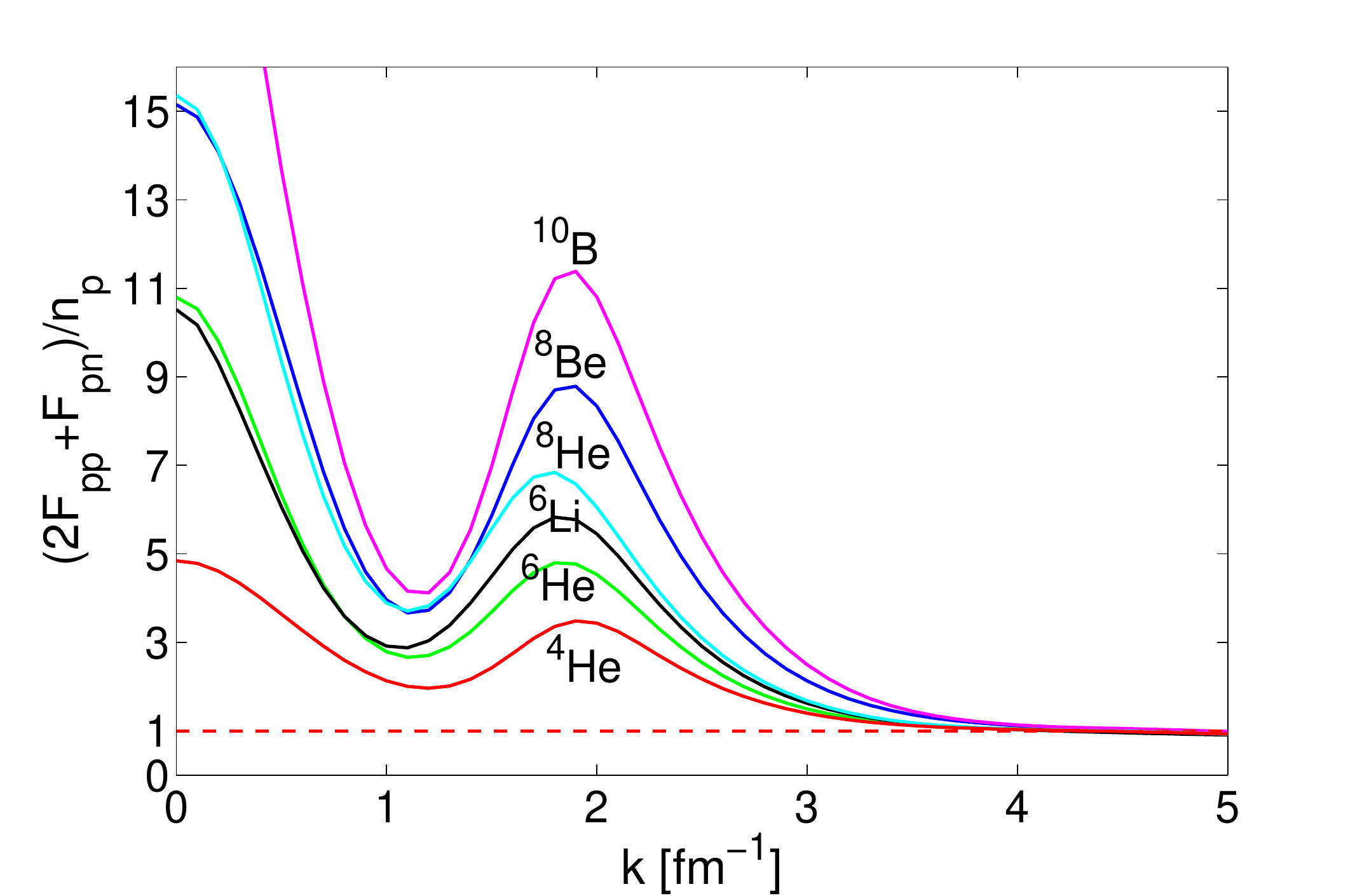}
\caption{\label{2VS1_protons} (Color online) 
The ratio $(2F_{pp}+F_{pn})/n_p$ for different nuclei.
The numerical data is taken from Ref. \cite{WirSchPie14}. Red 
line - $^4$He, green line - $^6$He, 
cyan line - $^8$He, black line - $^6$Li,
blue line - $^8$Be, and pink line - $^{10}$B.
The dashed red line is the reference $y=1$.}
\end{figure}

\begin{figure}
\includegraphics[width=8.6 cm]{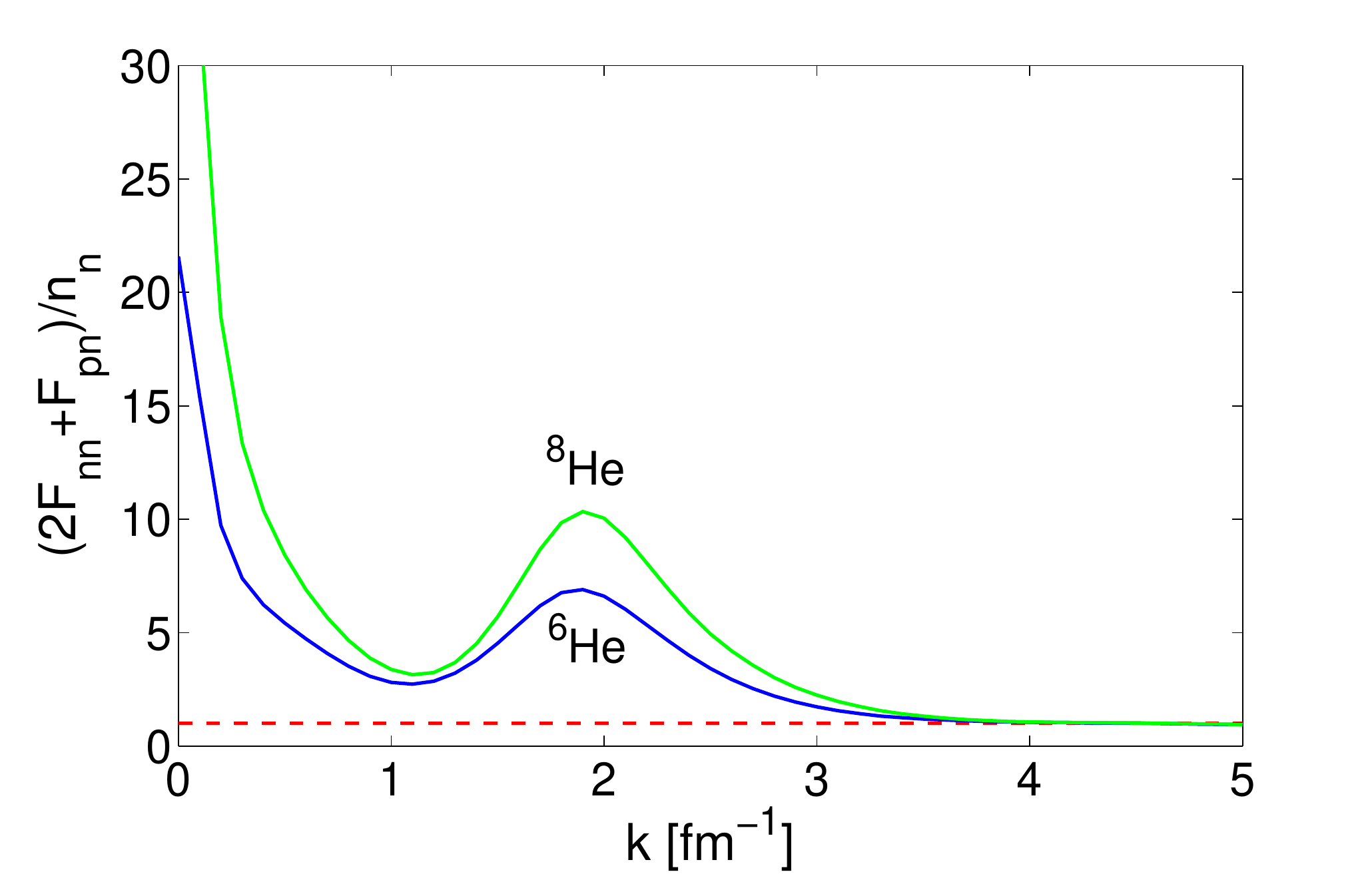}
\caption{\label{2VS1_neutrons} (Color online) 
The ratio $(2F_{nn}+F_{pn})/n_n$ for the non-symmetric nuclei
in the numerical data of Ref. \cite{WirSchPie14}. Blue 
line - $^6$He, and green line - $^8$He.
The dashed red line is the reference $y=1$.}
\end{figure}

First we check the relation between the one-nucleon and the two-nucleon 
momentum distributions, Eqs. (\ref{1pto2}), (\ref{1nto2}).
In Fig. \ref{2VS1_protons} the ratio between $2F_{pp}+F_{pn}$ and $n_p$ is 
presented for various nuclei.
We can see that for $k\longrightarrow\infty$ the two quantities
coincide and our prediction (\ref{1pto2})
is indeed satisfied. 
In Fig. \ref{2VS1_neutrons} we present the ratio
between $2F_{nn}+F_{pn}$ and $n_n$. We show
only the results for non-symmetric nucei, because for
symmetric nuclei
there is no difference between protons
and neutrons in the numerical VMC data.
We can see that also here, the ratio  $(2F_{nn}+F_{pn})/n_n \longrightarrow 1$
as  $k\longrightarrow\infty$ and our prediction (\ref{1nto2})
is satisfied.
This result is obtained for all available nuclei: $^4$He, $^6$He, $^8$He, 
$^6$Li, $^8$Be, and $^{10}$B, for both protons and neutrons. 
For all these nuclei the momentum relations hold for 
$4\:\mathrm{fm^{-1}} < k < 5\:\mathrm{fm^{-1}}$.

The correspondence between our predictions, derived using the 
contact formalism, and the numerical data is a good indication for the
relevance of the contact formalism to nuclear systems. We also learn here
that the approximations made in the above theoretical
derivations for $k\longrightarrow \infty$ are valid for 
$4\:\mathrm{fm^{-1}} < k < 5\:\mathrm{fm^{-1}}$. This is the first indication for the 
momentum range which is relevant to the contact formalism in nuclear systems.
Moreover, as we mentioned before, in current studies of 
SRCs in nuclei this momentum range of $k>4\:\mathrm{fm^{-1}}$
is believed to be affected by three-body correlations.
As explained, Eqs. (\ref{1pto2}) and (\ref{1nto2}) are 
suppose to be satisfied when the two-body correlations
are the only significant correlations and every high momentum
nucleon has a sinlge nucleon near it with back-to-back momentum.
It means that according to this numerical data the momentum range of 
$4\:\mathrm{fm^{-1}} < k < 5\:\mathrm{fm^{-1}}$ is affected almost exclusively
by two-body SRCs while three-body SRCs are negligible, and that in this
momentum range the picture of back-to-back short-range correlated 
pairs is accurate.

We note that this momentum range of $4\:\mathrm{fm^{-1}} < k < 5\:\mathrm{fm^{-1}}$
might be model dependent, and it should be
verified using other numerical methods, and different nuclear potentials. 
It should also be mentioned 
that the VMC method utilize 
two and three-body Jastrow correlations in
the nuclear wave function.

Hen et al. \cite{HenArx14} also discuss the possibility that the contact
formalism is relevant in nuclear physics. In their work,
they present an experimental measurement
of a $k^{-4}$ behavior in the proton momentum 
distribution in the deuteron for $1.6\:\mathrm{fm^{-1}}< k<3.2\:\mathrm{fm^{-1}}$.
They also claim that the $k^{-4}$ behavior exists in heavier nuclei
in the same momentum range.
As mentioned before, one of the results of the contact formalism in atomic systems
is the $k^{-4}$ tail in the momentum distribution, but this
behavior is a direct consequence of the zero-range model.
In nuclear systems this model is not accurate, so
we can only expect a high momentum tail universal to all nuclei,
but not a $k^{-4}$ behavior. We also note that in the numerical
VMC data there is no $k^{-4}$ tail for nuclei
heavier than the deuteron. Moreover, 
we have found here that the relevant momentum range for the
contact formalism in nuclei is
$4\:\mathrm{fm^{-1}} < k < 5\:\mathrm{fm^{-1}}$,
which is higher than the momentum range discussed by Hen et al.

\subsection{The $pp$ and $nn$ contacts along the nuclear chart}

\begin{figure}
\includegraphics[width=8.6 cm]{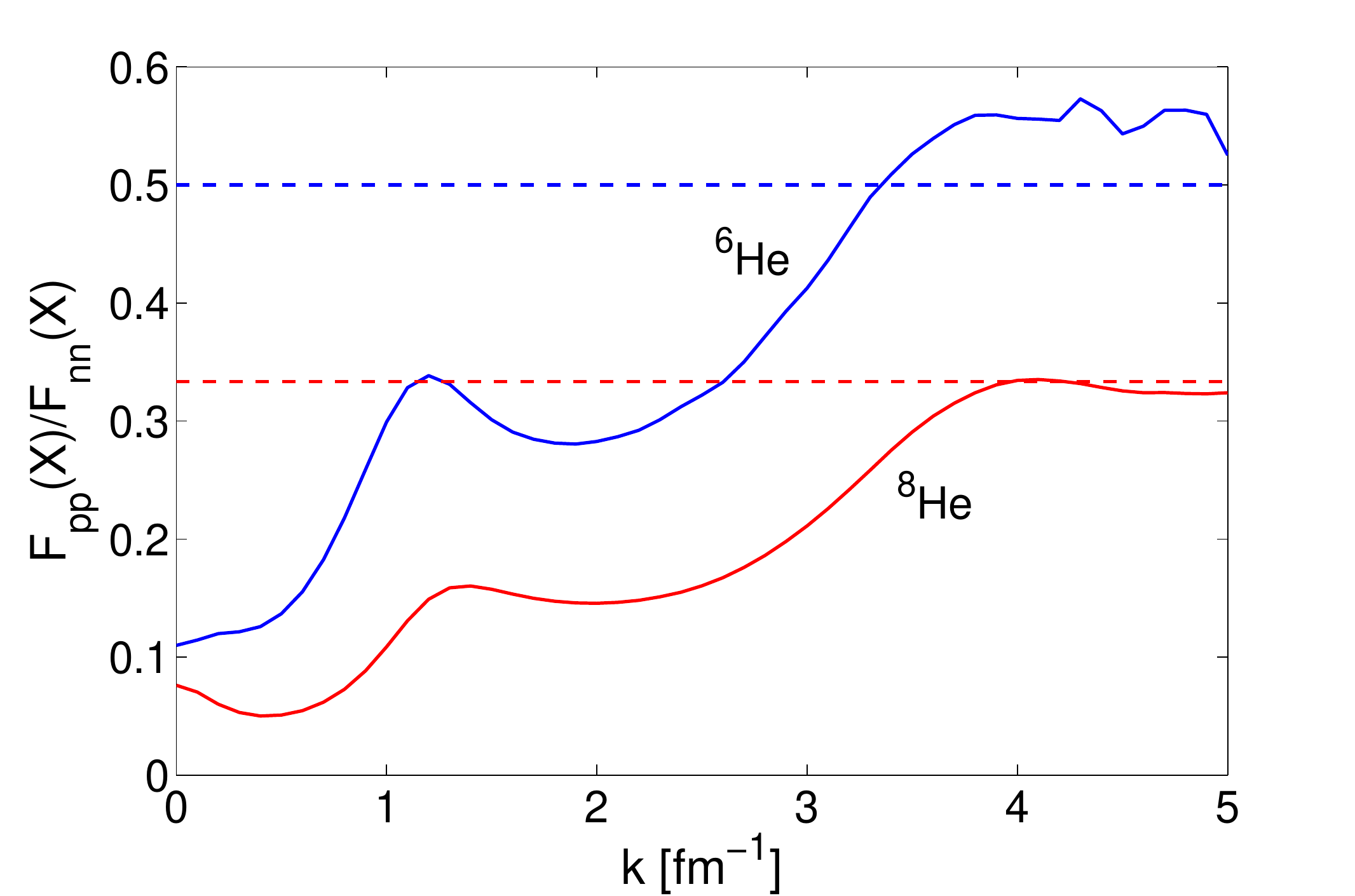}
\caption{\label{F_pp_F_nn} (Color online) 
The ratio between $F_{pp}$ and $F_{nn}$ in the same nuclei
 for the available non-symmetric nuclei in \cite{WirSchPie14}.
Blue line - $^6$He, and red line - $^8$He.
The dashed blue and red lines indicate the
value of $Z/N$ in $^6$He, and $^8$He, respectively.}
\end{figure}

We continue now by examining the 
ratio between $F_{pp}(\bs{k})$ and $F_{nn}(\bs{k})$ in 
the same nuclei. In the VMC results, this ratio equals 1
for all $k$ for symmetric nuclei $(N=Z)$. Therefore, we are
left with the available non-symmetric nuclei $^6$He and 
$^8$He. The relevant results are shown in Fig. \ref{F_pp_F_nn}.

We can see that for $4\:\mathrm{fm^{-1}} < k < 5\:\mathrm{fm^{-1}}$ the ratio is
approximately constant. Inspecting Eq. (\ref{2nuc}), we see that
the only way for this ratio to be constant is that {\it (i)} only
pairs in $\alpha,\beta$ states with the same $k$-dependence of
$\tilde{\varphi}_{ij}^{\alpha\dagger}\tilde{\varphi}_{ij}^{\beta}$
contribute significantly to 
both $F_{pp}$ and $F_{nn}$, and {\it(ii)} 
both $pp$ and $nn$ pairs have the same $k$-dependence.
It is reasonable to 
assume that the $s$-wave
contacts are the most significant contacts. 
For $pp$ and $nn$ pairs the only possible non-zero $s$-wave contact
is $C_{ij}^{\alpha_{00}\alpha_{00}}$ where $\alpha_{00}\equiv(s_2=0,\ell_2=0,j_2=0,m_2=0)$.
This point can be verified numerically through analysis of the angular dependence of the 
momentum distributions. If the $s$-wave contact is indeed dominant
we expect to see no angular-dependence. If we further assume that
$\tilde{\varphi}_{pp}^{\alpha_{00}\dagger}\tilde{\varphi}_{pp}^{\alpha_{00}}
=\tilde{\varphi}_{nn}^{\alpha_{00}\dagger}\tilde{\varphi}_{nn}^{\alpha_{00}}$,
which seems reasonable from isospin symmetry,
then the ratio between $F_{pp}$ and $F_{nn}$ for large momentum 
equals to the ratio between $C_{pp}^{\alpha_{00}\alpha_{00}}$
and $C_{nn}^{\alpha_{00}\alpha_{00}}$.

We can also see in  Fig. \ref{F_pp_F_nn}, that for the two relevant nuclei the 
ratio $F_{pp}/F_{nn}$ is close to the ratio $Z/N$ between the number of
protons and neutrons in the nucleus. 
If this relation turns out to be true in general along the nuclear chart, 
it means that for a nucleus $X$
in its ground state, the most significant
$pp$ and $nn$ contacts are 
$C_{pp}^{\alpha_{00}\alpha_{00}}$ and $C_{nn}^{\alpha_{00}\alpha_{00}}$
and their ratio is given by
\be \label{cpp2cnn}
   \frac{C_{pp}^{\alpha_{00}\alpha_{00}}(X)}{C_{nn}^{\alpha_{00}\alpha_{00}}(X)}
     \approx\frac{Z(X)}{N(X)},
\ee
and
\be
\varphi^{\alpha_{00}\alpha_{00}}_{pp}(r)=\varphi^{\alpha_{00}\alpha_{00}}_{nn}(r).
\ee
Here $Z(X)$ ($N(X)$) is the number of protons (neutrons) in the nucleus $X$.
This result is surprising because one might think that the ratio (\ref{cpp2cnn})
should scale as the ratio between the total number of $pp$
pairs and the number of $nn$ pairs in the nucleus, i.e. $Z^2/N^2$. The above result
tells us that the number of correlated $pp$ and $nn$ pairs in nuclei goes
like $Z$ and $N$, respectively.
If we check the ratio between $F_{pp}$ or $F_{nn}$ and $F_{pn}$,
no plateau is observed.

\begin{figure}
\includegraphics[width=8.6 cm]{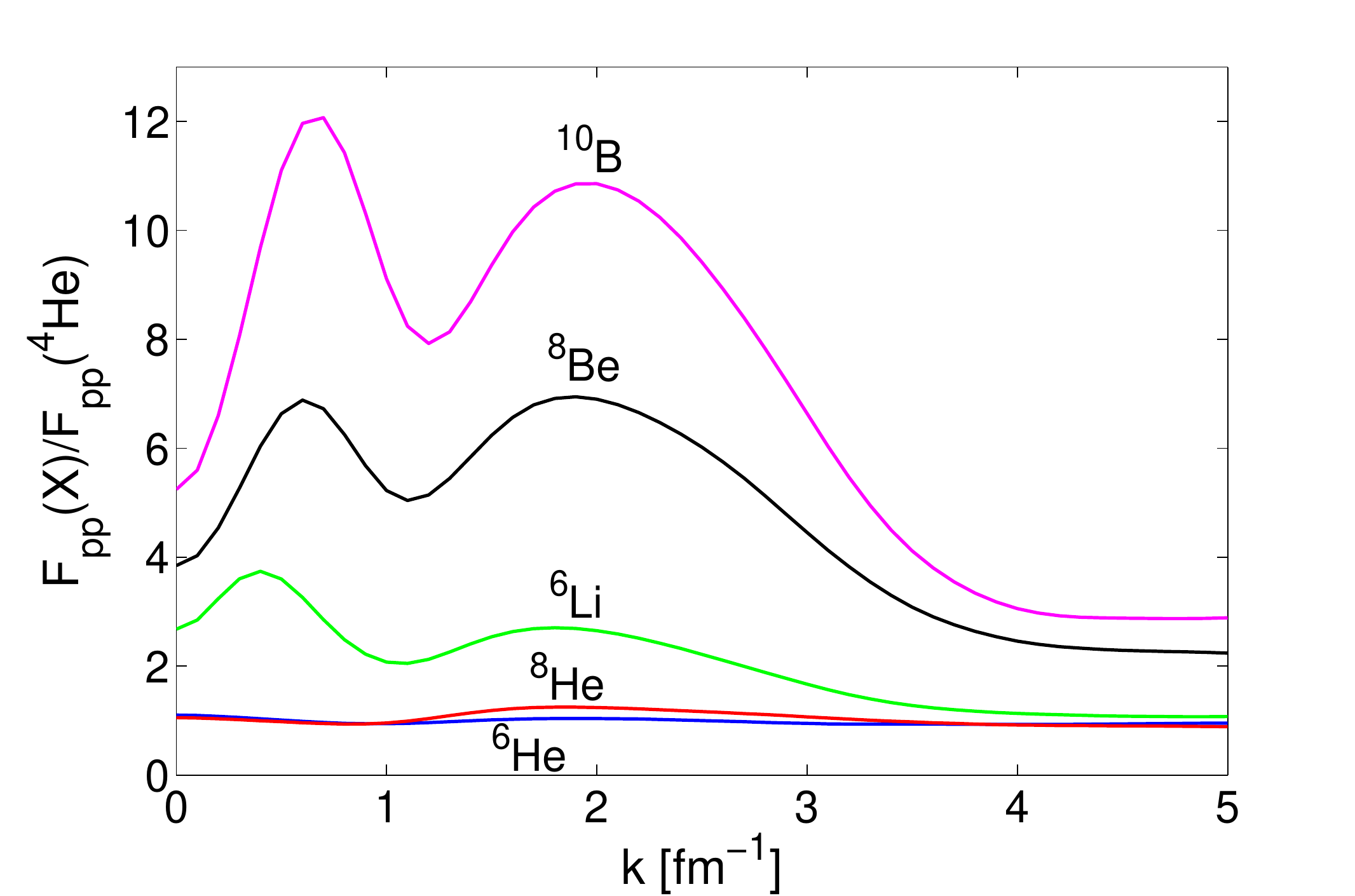}
\caption{\label{F_pp_F_pp} (Color online)
The ratio between  $F_{pp}(X)$ and $F_{pp}(^4\mathrm{He})$
 for the available nuclei X in the numerical data of Ref. \cite{WirSchPie14}.
Blue line - $^6$He, red line - $^8$He,
green line - $^6$Li, black line - $^8$Be,
and pink line - $^{10}$B.}
\end{figure}

We can also examine the ratio between
$F_{pp}$ of nucleus $X$ and $F_{pp}$ of another nucleus $Y$.
The results are presented in Fig. \ref{F_pp_F_pp}, where all
the available nuclei are compared to $^4$He. 

Here again we see flattening
for $4\:\mathrm{fm^{-1}} < k < 5\:\mathrm{fm^{-1}}$. This behavior supports
the claim that only one contact contributes
significantly to $F_{pp}$, and so the value of this ratio is just
the value of the ratio of this $pp$ contact in the two different nuclei
(see Eq. (\ref{2nuc})). The constant behavior also
supports the assumption that the pair wave functions 
$\varphi_{pp}^{\alpha\beta}$ are universal along the
nuclear chart, because that way the $k$-dependence indeed vanishes.
The average values of this ratio for $4\:\mathrm{fm^{-1}}\le k \le 5 \:\mathrm{fm^{-1}}$
are presented in table \ref{table_F_ppX_F_pp_4He} and compared to the ratio
between the number of protons in the relevant nuclei and the
number of protons in $^4$He. We can see that the two ratios
are approximately equal for the different nuclei.
If the most significant $pp$ contact is the $s$-wave
contact $C_{pp}^{\alpha_{00}\alpha_{00}}$, then we  
we can deduce that for nuclei $X$ and $Y$ in their ground state:
\be
\frac{C_{pp}^{\alpha_{00}\alpha_{00}}(X)}{C_{pp}^{\alpha_{00}\alpha_{00}}(Y)}\approx\frac{Z(X)}{Z(Y)}.
\ee
For $F_{nn}$ similar results are observed, therefore we can also deduce that
\be
\frac{C_{nn}^{\alpha_{00}\alpha_{00}}(X)}{C_{nn}^{\alpha_{00}\alpha_{00}}(Y)}\approx\frac{N(X)}{N(Y)}.
\ee
These relations support the claim that the number of 
correlated $pp$ and $nn$ pairs in nuclei is proportional to
$Z$ and $N$, respectively.

\begin{table}
\begin{tabular}{ c  c  c  }
  \hline
\hline                       
  $X$ & $\bra F_{pp}(X)/F_{pp}(^4\mathrm{He})\ket$ & $Z(X)/Z(^4\mathrm{He})$ \\
\hline
  $^6$He & $0.94\pm 0.01 (1\sigma)$ & 1 \\
  $^8$He & $0.90\pm 0.01 (1\sigma)$ & 1 \\
  $^6$Li & $1.09\pm 0.02 (1\sigma)$ & 1.5 \\
  $^8$Be & $2.31\pm 0.07 (1\sigma)$ & 2 \\
  $^{10}$B & $2.91\pm 0.06 (1\sigma)$ & 2.5 \\
\hline  
\hline    
\end{tabular}
 \caption{\label{table_F_ppX_F_pp_4He}The averaged value of the ratio
between $F_{pp}(X)$ and $F_{pp}(^4\mathrm{He})$ for
 $4\:\mathrm{fm^{-1}}\le k \le 5\:\mathrm{fm^{-1}}$ for all the available 
nuclei in the numerical data of Ref. \cite{WirSchPie14}.}
\end{table}

\subsection{The $pn$ contacts and the Levinger constant}

\begin{figure}
\includegraphics[width=8.6 cm]{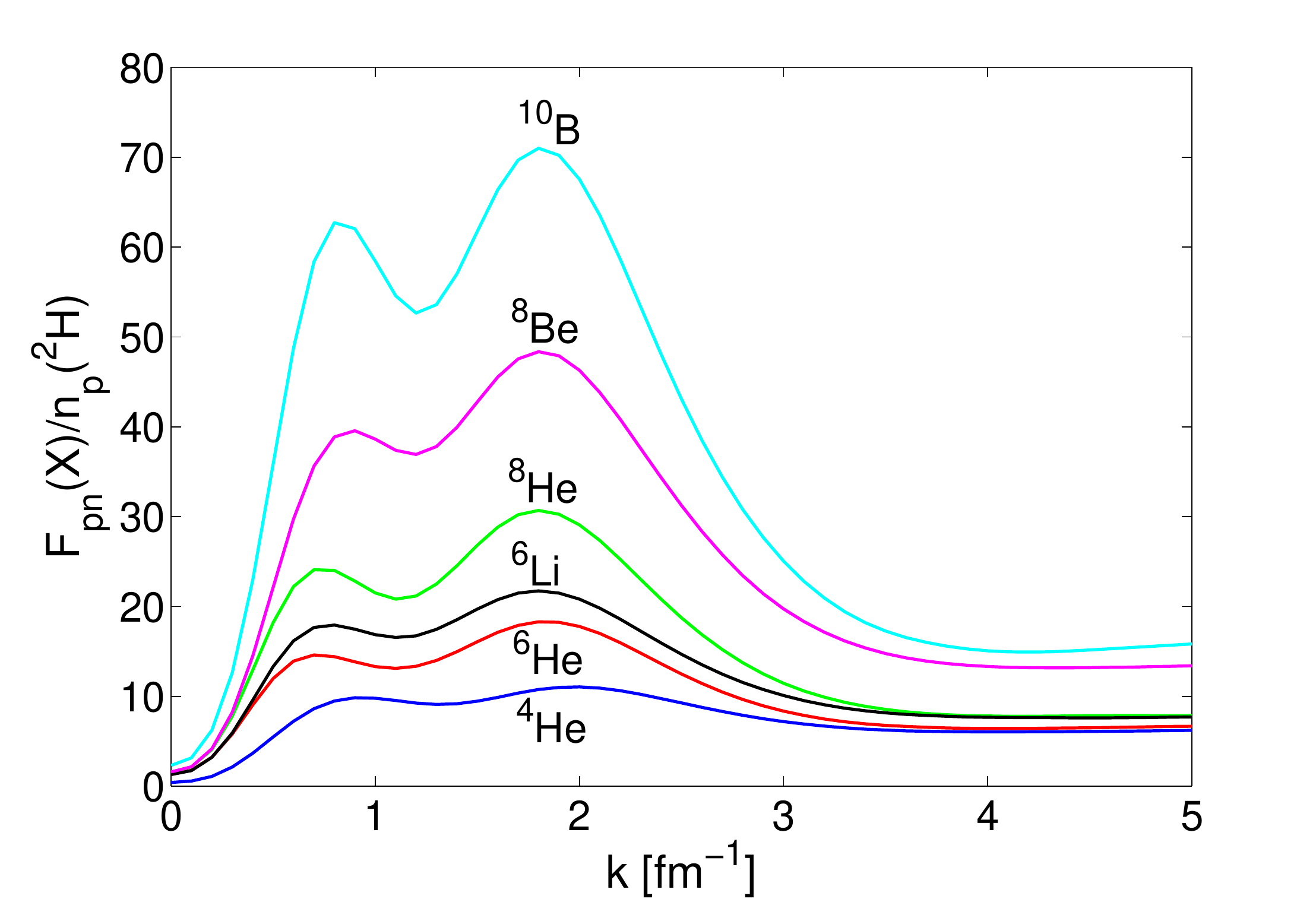}
\caption{\label{F_pn_n_p_2H} (Color online) 
 The ratio between $F_{pn}(X)$ and $n_p(^2\mathrm{H})$
 for the available nuclei X in the numerical data of Ref. \cite{WirSchPie14}.
Blue line - $^4$He, red line - $^6$He,
green line - $^8$He, black line - $^6$Li,
pink line - $^{8}$Be, and cyan line - $^{10}$B.}
\end{figure}

So far we have studied the properties of the $pp$ and $nn$
contacts, now we turn to study the $pn$
contacts. The $pn$ contacts might be the most interesting ones
because of the dominance of correlated $pn$ pairs in nuclear SRCs \cite{HenSci14}.
In order to study the properties of the $pn$ contacts we
examine the variation in $F_{pn}$ between different nuclei.
As in the $pp$ and $nn$ cases, also in this case we shall
assume that the $s$-wave is the most 
dominant partial wave. For a $pn$
pair  in an $s$-wave there are two possible spin configurations,
spin-singlet and spin-triplet.
For the deuteron $^2$H,
only the spin triplet is relevant as it is a $J=1$ state.
In  Fig. \ref{F_pn_n_p_2H}
we present the ratio between $F_{pn}(X)$ and $n_p(^2\mathrm{H})$
for the available nuclei in the VMC results. 

Once again, a constant behavior 
is seen for $4\:\mathrm{fm^{-1}} < k < 5\:\mathrm{fm^{-1}}$.
As mentioned before, we have three equal spin triplet $s$-wave $np$
contacts, $C_{pn}^{\alpha_{1\mu}\alpha_{1\mu}}$, and one spin-singlet
$s$-wave $np$ contact $C_{pn}^{\alpha_{00}\alpha_{00}}$. Moreover,
$|\tilde{\varphi}_{pn}^{\alpha_{1\mu}}|^2$
is independent of $\mu$.
Consequently, we would expect to see a plateau in the
ratio $F_{pn}(X)/n_p(^2\mathrm{H})$,
if either the asymptotic pair wave functions obey the relation
$|\tilde{\varphi}_{pn}^{\alpha_{1\mu}}|^2=
|\tilde{\varphi}_{pn}^{\alpha_{00}}|^2$
or alternatively if the spin-triplet $s$-wave contacts are dominant.
In the first case we can deduce from the relations between
the contacts and the one-nucleon and two-nucleon momentum 
distributions that asymptotically
\begin{align}
\frac{F_{pn}(X)}{n_p(^2\mathrm{H})} &\approx 
\frac{3 C_{pn}^{\alpha_{10}\alpha_{10}}(X)
          +C_{pn}^{\alpha_{00}\alpha_{00}}(X)}
     {3 C_{pn}^{\alpha_{10}\alpha_{10}}(^2\mathrm{H})} \nonumber \\
& = 
\frac{C_{pn}^{s_2=0}(X)+C_{pn}^{s_2=1}(X)}{C_{pn}^{s_2=1}(^2\mathrm{H})},
\end{align}
where here we have also used the notation
of Eqs. (\ref{rel_singlet}) and (\ref{rel_triplet}).
In the second case we get
\be
\frac{F_{pn}(X)}{n_p(^2H)}
  \approx
     \frac{C_{pn}^{\alpha_{10}\alpha_{10}}(X)}
          {C_{pn}^{\alpha_{10}\alpha_{10}}(^2\mathrm{H})} 
  =
     \frac{C_{pn}^{s_2=1}(X)}{C_{pn}^{s_2=1}(^2H)}.
\ee
In a previous paper \cite{WeiBazBar14}, we have predicted
that the ratio between the sum of the two $s$-wave $np$ contacts
of a nucleus $X$ in his ground state
and the deuteron's $s$-wave $np$ contact is given by
\be
\frac{C_{pn}^{s_2=0}(X)+C_{pn}^{s_2=1}(X) }{C_{pn}^{s_2=1}(^2\mathrm{H})}=L\frac{NZ}{A},
\ee
where $L$ is Levinger's constant that
relates, at the high energy hand, the photoabsorption
cross section of a nucleus to the photoabsorption cross section
of the deuteron \cite{Lev51}.
Analysis of the experimental results \cite{TavTer92}
suggest that the $L$ is approximately a constant
along the nuclear chart $L\approx5.50\pm 0.21$ \cite{WeiBazBar14}.
In \cite{WeiBazBar14}, we have assumed that the two $s$-wave
states have the same asymptotic pair wave function in small distances,
which corresponds to the first case above.
If we were to assume that only the spin-triplet $np$ $s$-wave
is significant, then our result would have been
\be
\frac{C_{pn}^{s_2=1}(X)} {C_{pn}^{s_2=1}(^2\mathrm{H})}=L\frac{NZ}{A}.
\ee
In any of the two cases, we get the relation
\be
\frac{F_{pn}(X)}{n_p(^2\mathrm{H})}\approx
L\frac{NZ}{A},
\ee
that should hold in the high momentum range.
For this range of high momentum the ratio between
$F_{pn}$ and $n_p(^2\mathrm{H})$ is the number of 
quasideuteron (qd) pairs with high relative momentum
in the nucleus.
In table \ref{table_F_pnX_n_p_2H} we present the 
averaged value of this ratio for $4\:\mathrm{fm^{-1}}\le k \le 5\:\mathrm{fm^{-1}}$
and its multiplication by $A/NZ$ for each nuclei $X$,
which should be equal to $L$ according to the above
prediction.
One can see that the values of the multiplied ratio
are close to the above value of $L$ for all the nuclei
and their average value is $5.7\pm 0.7(1\sigma)$.
This value is in a very good agreement
with the above mentioned value of $L$.

Evaluation of Levinger's constant from the number
of qd pairs was done by Benhar et al. \cite{BenFabFan03}.
In their work, they calculate numerically the number of qd
pairs in the nucleus and extract Levinger's constant.
In our evaluation we consider only the qd pairs
with high relative momentum, which corresponds to
small relative distance. Only such qd pairs can be
emitted in the photoabsorption process, and therefore
only  they should be considered.

We have compared here two independent
relations between the $np$ contacts and different 
properties of nuclei (momentum distribution and 
photoabsorption cross section) and obtained a good agreement 
between the two. Doing so, we have also obtained here an
established estimation for the leading
$s$-wave $np$ contact(s) along the nuclear chart
for nuclei in their ground state
(in units of the deuteron's $s$-wave $np$ contact).

\begin{table}
\begin{tabular}{c  c  c }
\hline  
\hline                       
  $X$ & $\bra F_{pn}(X)/n_{p}(^2H)\ket$ & $A/NZ\bra F_{pn}(X)/n_{p}(^2H)\ket$ \\
  \hline
  $^4$He & $6.10\pm0.06 (1\sigma)$ &$6.10\pm0.06$  \\
  $^6$He & $6.5\pm 0.1 (1\sigma)$ & $4.88\pm 0.08$ \\
  $^8$He & $7.82\pm 0.03 (1\sigma)$ & $5.21\pm 0.02$ \\
  $^6$Li & $7.63\pm 0.04(1\sigma)$ &$5.09\pm 0.03$ \\
  $^8$Be & $13.25\pm0.08(1\sigma)$ &$6.63\pm0.04$  \\
  $^{10}$B & $15.3\pm 0.3(1\sigma)$ &$6.1\pm 0.1$ \\
  \hline    
\hline
\end{tabular}
 \caption{\label{table_F_pnX_n_p_2H}The 
averaged value of the ratio
between $F_{pn}(X)$ and $n_p(^2\mathrm{H})$ for
 $4\:\mathrm{fm^{-1}}\le k \le 5\:\mathrm{fm^{-1}}$ and its
multiplication by $A/NZ$ for all the available 
nuclei in the numerical VMC results of \cite{WirSchPie14}.}
\end{table}

\section{Summary} 
Summing up, 
we have generalized the contact formalism to nuclear systems and defined
a matrix of contacts for each particle pair: $pp$, $nn$ and $pn$. With this 
generalization we have taken into consideration both different 
partial waves, and finite-range interaction.
We have discussed the simple properties of the nuclear contacts and demonstrated
the use of the generalized formalism by relating the contacts to the one-nucleon
and two-nucleon momentum distributions. As a result we have obtained a relation
between these two momentum distributions, which emphasizes the significant 
contribution of SRCs to the high one-nucleon momentum tail. 
Using avilable VMC numerical data \cite{WirSchPie14}, we have verified the above 
relation and found further relations 
between the different nuclear contacts.
Using few of these new relations and a previous prediction
connecting the $pn$ contacts to the Levinger constant, we 
have calculated Levinger's constant for the available nuclei
and got a good agreement with its experimental value.
This is an important indication 
for the relevance of the contact formalism to nuclear systems, and might 
open the path to revealing many more interesting relations.
We have also learned from the numerical data that the relevant
 momentum range for the contact's approximations in nuclear systems
is $4\:\mathrm{fm^{-1}} < k < 5\:\mathrm{fm^{-1}}$. However, we note that this result might be model
dependent.
The fact that the relations between the one-nucleon
and two-nucleon momentum distribution were satisfied
in this momentum range teaches us that for such momenta
the two-body SRCs, rather than three-body SRCs, are dominant.
Additional numerical or experimental data for both
one-nucleon and two-nucleon momentum distributions in more nuclei,
also in excited states, including angular-dependence
is needed in order to improve our understanding regarding the
properties of the nuclear contacts.

\begin{acknowledgments}
This work was supported by the Pazy foundation.
\end{acknowledgments}


\begin{thebibliography}{99}
\bibitem{Tan08}%
  S. Tan, Ann. Phys. (N.Y.) {\bf 323}, 2952 (2008); 
          {\bf 323}, 2971 (2008); {\bf 323}, 2987 (2008). 
\bibitem{Bra12}%
  E. Braaten, in {\it BCS-BEC Crossover and the Unitary Fermi Gas}, edited by 
  W. Zwerger (Springer, 2012)
\bibitem{SteGaeDra10}%
  J. T. Stewart, J. P. Gaebler, T. E. Drake, and D. S. Jin, 
  Phys. Rev. Lett. {\bf 104}, 235301 (2010).
\bibitem{SagDraPau12}%
  Y. Sagi, T. E. Drake, R. Paudel, and D. S. Jin,
  Phys. Rev. Lett. {\bf 109}, 220402 (2012).
\bibitem{ParStrKam05}%
  G. B. Partridge, K. E. Strecker, R. I. Kamar, M. W. Jack, and R. G. Hulet,
  Phys. Rev. Lett. {\bf 95}, 020404 (2005).
\bibitem{WerTarCas09}%
  F. Werner, L. Tarruel, and Y. Castin, Eur. Phys. J. B {\bf 68}, 401 (2009).
\bibitem{KuhHuLiu10}%
  E. D. Kuhnle, H. Hu, X.-J. Liu, P. Dyke, M. Mark, P. D. Drummond, 
  P. Hannaford, and C. J. Vale, 
  Phys. Rev. Lett. {\bf 105}, 070402 (2010).
\bibitem{WirSchPie14}%
  R. B. Wiringa, R. Schiavilla, S. C. Pieper, J. Carlson, 
  Phys. Rev. C {\bf 89}, 024305 (2014).
\bibitem{Fmoin12}%
  N. Fomin et al., 
  Phys. Rev. Lett. {\bf 108}, 092502 (2012).
\bibitem{HenSci14}%
  O. Hen et al. (CLAS Collaboration), 
  Science {\bf 346}, 614 (2014).
\bibitem{AlvCioMor08}%
  M. Alvioli, C. Ciofi degli Atti and H.morita, 
  Phys. Rev. Lett. {\bf 100}, 162503 (2008).
\bibitem{ArrHigRos12}%
  J. Arrington, D. Higinbotham, G. Rosner, M. Sargsian, 
  Prog. Part. Nucl. Phys. {\bf 67}, 898 (2012).
\bibitem{Egiyan06}%
  K. Egiyan et al. (CLAS Collaboration), 
  Phys. Rev. Lett. {\bf 96}, 082501 (2006).
\bibitem{WeiBazBar14}
  R. Weiss, B. Bazak, and N. Barnea, 
  Phys. Rev. Lett. {\bf 114}, 012501 (2015).
\bibitem{WerCas12}%
  F. Werner and Y. Castin, 
  Phys. Rev. A {\bf 86}, 013626 (2012).
\bibitem{zerorange}%
  H. A. Bethe and R. Peierls, 
  Proc. Roy. Soc. {\bf 148}, 146 (1935).
\bibitem{BraKanPla11} 
  E. Braaten, D. Kang, and L. Platter,
  Phys. Rev. Lett. {\bf 106}, 153005 (2011).
\bibitem{HenArx14}%
  O. Hen, L.B. Weinstein, E. Piasetzky, G.A. Miller, M.M. Sargsian and Y. Sagi, 
  arXiv:1407.8175v2 [nucl-ex] (2014).
\bibitem{Lev51}%
  J. S. Levinger, 
  Phys. Rev. {\bf 84}, 43 (1951).
\bibitem{TavTer92}%
  M. L. Terranova, D. A. De Lima and J. D. Pinheiro Filho,
  Europhys. Lett. {\bf 9} 523 (1989); 
  O. A. P. Tavares and M. L. Terranova, J. Phys. G {\bf 18}, 521 (1992).
\bibitem{BenFabFan03}%
  O. Benhar, A. Fabrocini, S. Fantoni, A. Yu. Illarinov, G. I. Lykasov,
  Phys. Rev. C {\bf 67}, 014326 (2003).
\end{thebibliography}
\end{document}